\documentclass[dvips,11pt]{article}

\usepackage[latin1]{inputenc}  

\usepackage[american]{babel}
\usepackage{amsmath,amssymb,amsfonts,amsthm,mathrsfs}

  \usepackage{graphics} 
  \usepackage{epsfig} 
\usepackage{graphicx}  \usepackage{epstopdf}

\theoremstyle{remark}\newtheorem{remark}{Remark}

\graphicspath{{./}{figure/}}

\begin{document}

{\bf \title {Escaping the trap of 'blocking': a kinetic model linking economic development and political competition. }}

\author{Marina Dolfin\thanks{Department of   Engineering,  University of Messina, Messina, Italy}
\and Dami\'an Knopoff\thanks{Centro de Investigaci\'on y Estudios de Matem\'atica (CONICET) -- FaMAF (UNC),C\'ordoba, Argentina )} \and Leone Leonida \thanks{Department of Management, King's College, London, UK} \and Dario Maimone Ansaldo Patti \thanks{Department of Economics, University of Messina, Messina, Italy }}

\date{}
\maketitle

\begin{abstract}
In this paper we present a kinetic model with stochastic game-type interactions, analyzing the relationship between the level of political competition in a society and the degree of economic liberalization.
The above issue regards the complex interactions between economy and institutional policies intended to introduce technological innovations in a society, where technological innovations are intended in a broad sense comprehending reforms critical to production  \cite{AR2015}. A special focus  is placed  on the {\it political replacement effect} described in a macroscopic model by Acemoglu and Robinson (AR-model \cite{acemoglu}, henceforth), which can determine the phenomenon of innovation 'blocking', possibly  leading to economic backwardness.   One of the goals of our modelization is to obtain a mesoscopic dynamical  model whose macroscopic outputs  are qualitatively comparable with  stylized facts of the AR-model.      A set of numerical solutions is presented showing the non monotonous  relationship between economic liberization and political competition, which can be considered as an emergent phenomenon of the complex socio-economic interaction dynamic. 

\end{abstract}

\section{Introduction}

Understanding the differences among the rates of industrialization and the introduction and spreading of  technological innovation in different countries, possibly leading to economic backwardness in some of them, became a central issue in economic studies starting from the seminal essay by Gerschenkron \cite{gersch}.  In this context a particular focus has been applied on the role played by  political elites, although the relationship between political and economic perspectives appears still controversial. An interesting hypothesis on the interplay between political perspectives and possibly economic backwardness  has been proposed in some papers by   Acemoglu and   Robinson (see \cite{acemoglu, AR2000} and references therein). Their model proposes as testing hypothesis a nonmonotonous relationship between the introduction of technological  innovation by the incumbent ruler and the level of political competition in the society. This hypothesis has been statistically tested with positive results  by two of the authors \cite{leonida} using data on 102 countries over the period 1980 to 2005.

Our project consists in developing a model in the framework of the Kinetic Theory for Active Particles (KTAP) \cite{[BKS13]}, which would allow to recover    stylized facts of the model proposed by Acemoglu and Robinson although  in a different mathematical setting and in dynamical conditions.
   In fact, 
the trap cited in our paper's  title refers to the fact that in the AR-model, when the incumbent rulers find it more convenient to block the introduction of new technologies in the society,  the same will happen again and again in the future. Our intuitive idea is that our   dynamical modelization of the same phenomenology may find conditions on the model parameters avoiding this kind of 'trap'.

 More precisely, the paper analyses the interplay among the introduction of technological innovation by an incumbent ruler in a society leading to citizen income increment, the  political support/opposition of the citizens  and the political competition.
We cluster the  population into three groups (functional subsystems in KTAP), each population within a group being homogeneously distributed, i.e. no dependance on the space variable is assumed   and the socio-political determinants of each group are represented by two  socio-economic variables.  The interactions are modeled using a stochastic game-type approach.

The contents of this paper refers also  to a recent contribution regarding a kinetic model with a bivariate distribution and concerning the modeling of the interaction of welfare policy and support/opposition to governments
\cite{[BHT13]}, where it is shown that the interactions of two different dynamics can lead to radicalization of the opposition.
A detailed analysis of the role of nonlinear interactions is proposed in \cite{[DL14]}, where it is shown how the overall wealth of a nation can be influenced by different models of social interactions.

The paper is organized as follows: in Sec. 2 we present the phenomenology of the system that we are going to model  and that is based on a complex dynamic  between economy and policy inspired  by a model by Acemoglu and Robinson\cite{acemoglu, AR2000};  in Sec. 3 we introduce the mathematical representation of a society partitioned into three interacting subsystems such that on each subsystem two different dynamics take place and without considering migration among the subsystems. Then, we introduce the specific case modeling the  complex outcomes of the interacting phenomena of economic development and political policies among three specific groups of interest, individuated as {\it ruler}, {\it citizens} and a political {\it competing group}, focussing on the comparisons between the AR-model and the presented kinetic model. In Sec. 4 we present a set of numerical solutions   in order to test AR-model hypothesis with the outcomes of the proposed kinetic model. Moreover, we propose other possible scenarios.

 \section{Analysing the role of political losers in strategies of economic development}\label{modello}

\subsection{Phenomenology description toward a modeling strategy}

Political institutions have a direct influence on the economic development of a society, by means of general economic incentives and reforms. Technological innovation is here intended in a broad sense, as clarifyed by Acemoglu and Robinson: ''enforcement of property rights such as the creation of new legal institutions or the removal of regulations that prevent productive activities" \cite{AR2000}. As already remarked, an important issue regards the dynamics that possibly leads to the phenomenon of  'blocking' of economic incentives and then to economic backwardness in the society.
An hypothesis   is that the blocking of technological development is more due to the  fear of losing  political power that to fear of losing  economic rents   and the major point in analyzing 'blocking'  is in this case   the threat that innovation poses to the political power more than to economic rents \cite{AR2000}.
On this path, Acemoglu and Robinson suggest that the greatest impediment to economic development comes not from groups whose economic interests are adversely affected  by economic changes, but from elites whose political power is threatened; basically from political losers instead than from economic losers.
Then, the effect of economic changes on political power is a key factor in determining whether technological advances and beneficial economic changes will be blocked by groups whose political power, more than economic rents, is eroded. This observation suggests to look more to the determinants of the distribution of political power in view of understanding the 'blocking' conditions. The Austrian-Hungarian historical situation is taken as an example of the so called 'political replacement effect' \cite{acemoglu}; one may have a   significant view of it in the book 'Der Mann ohne Eigenschaften' ('The man without qualities') by   Musil \cite{musil}; moreover in Musil's book we  find discussions about the possibility of analyzing collective phenomena   borrowing ideas from statistical mechanics.

In AR-model three groups are considered, characterized by different averaged quantities: the incumbent ruler, the citizens and a political competing group. Summarizing, the model focuses on the conditions determining the blocking of technological innovations, understood in a broad sense, by means of the ruler; the emerging behavior is represented by the political replacement effect, i.e. the innovation blocking by means of the political elites, determined by the fear to lose power. In the following we introduce the basic parameters of the AR-model.

\subsection{Parameters of AR-model and  'blocking' condition }
 
AR-model is based on the  observation that innovation  induces 'turbulence'  which may erode the ruler's power.
The complex dynamics involving the introduction  of technological innovation by an incumbent ruler  and political competition is exploited making use of three parameters:

\begin{itemize}
\item $\mu$ -  inverse  measure of the level of   political competition in the society.

\item $\alpha$ - quantifying  the effect of the introduction of technological innovation by the ruler on the production. 

\item
$\gamma$ -  quantifying  the 'erosion' of the political power of the incumbent rulers due to the introduction of new technologies in the society. 

\end{itemize}

The 'blocking' condition is verified when the payoff of the ruler in the case in which they do not introduce innovations is bigger than the analogous value calculated when they    innovate. The 'blocking' condition is  obtained in \cite{acemoglu} as

\begin{equation}\label{cond2}
\alpha\,P\Big[\frac 1{2}+\mu\Big ]  \geq P\Big [\frac 1{2} +\gamma\,\mu - (\alpha -1)\Big ] \end{equation}
with
$$\footnotesize{
P[h]=
\left\{\begin{array}{l}
\displaystyle{0\hspace{1cm}\mathrm{if}\qquad h<0}\\\\
\displaystyle{h\hspace{1cm}\mathrm{if}\qquad 0\leq h \leq 1}
\\\\ \displaystyle{1 \hspace{1cm}\mathrm{if}\qquad h >1.}
\end{array} \right.}
$$

A central result is the nonmonotonous relationship between the propensity of the ruler to innovate and the political competition in the society, in a particular range of parameter values. In particular, the authors claim that the ruler may be induced to not innovate whenever the political competition in the society takes values in a medium range. For high or low values of political competition the ruler would always decide to innovate; this result will be quantified in the following.
We 
define  a function $\mathcal F(\mu,\,\gamma;\,\alpha)$  characterizing the introduction of technological innovations by the ruler, called {\it innovation function}:

\begin{equation}\label{inno}
\mathcal F(\mu,\,\gamma;\,\alpha)= \alpha\,P\Big[\frac 1{2}+\mu\Big ] -  P\Big [\frac 1{2} +\gamma\,\mu - (\alpha -1)\Big ],
\end{equation}
with

\begin{equation}\label{inno1}
\mathcal F(\mu,\,\gamma;\,\alpha)\::\: (0,\infty)\, \times\, (1,\infty)\to \mathbb R;
\end{equation}
i.e. it is a function depending on the  political competition characterized by $\mu$ and the  'turbulence' due to innovation $\gamma$; moreover it is parametrized by $\alpha$.
 It is clear from \eqref{cond2} that  the innovation  'blocking' phenomenon   appears in the range of values for the variables and the parameter, in which the innovation function is negative; in fact it is in this range of values that  it is more convenient for the incumbent ruler to 'block' the introduction of technological innovation  with respect to introducing new technologies, comparing the respective payoffs.
 In the following we will discuss numerically some of these significative values.

\subsection{Nonmonotonic relationship between innovation and political competition}
 In Fig. $1_a$, the innovation function $\mathcal F$ is plotted  for $\alpha=1.1$. The 'blocking region', defined as the domain of the variables such that $\mathcal F<0$,  is underlined in  Fig. $1_b$ plotting $\mathcal F$  for a the same value of  $\alpha$ and for  $\gamma =2$.

\begin{figure}[htbp]
 \vspace{-.2cm}
\centering{\includegraphics[width=13cm]{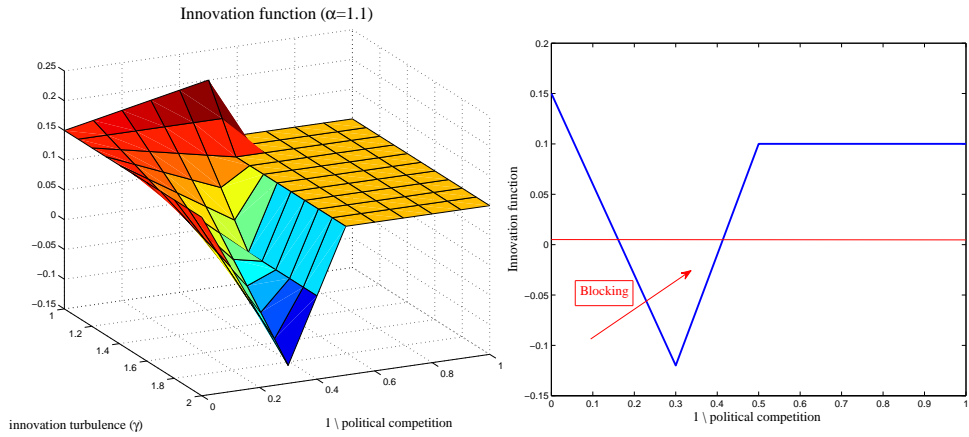}}
 \caption{{\bf Nonmonotonicity with 'blocking'}.}
\end{figure}

Moreover, we plotted the innovation function for  $\alpha=1.167$ (Fig.$2_a$)  and $\alpha=3.5$ (Fig.$3_a$)   and the corresponding plots obtained fixing $\gamma=2$ (Fig.$2_b$ and Fig.$3_b$, respectively).
\begin{figure}[htbp]
\centering{\includegraphics[width=13cm]{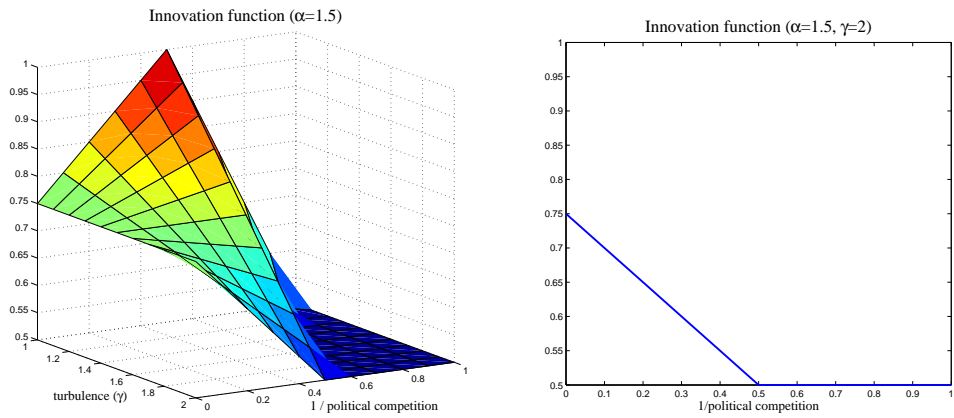}}
 \caption{{\bf Nonmonotonicity without 'blocking'}.} 
 
\end{figure}

\begin{figure}[htbp]
 \vspace{1cm}
\centering{\includegraphics[width=13cm]{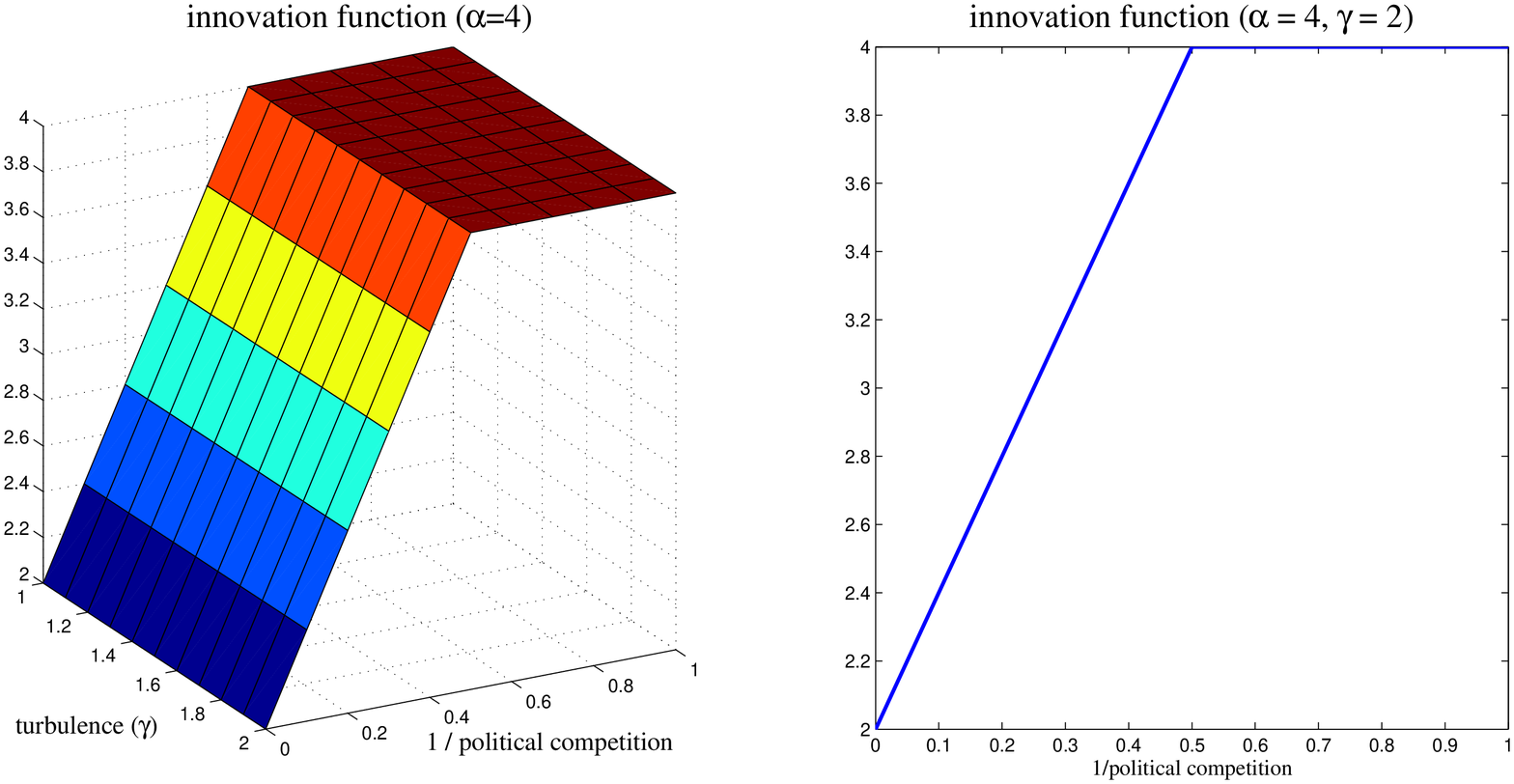}}
 \caption{{\bf Linearity without 'blocking'}.} 
\end{figure}

One  observes that the blocking condition is verified only for values of $\alpha$ 
near to 1; moreover when $\alpha$ grows the innovation as function of the political competion loses its nonmonotonicity becoming linear, as one can observe from Fig. 3. In particular,  $\alpha=1.167$ corresponds to the minimal parameter value such that the innovation function doesn't attain negative values and $\alpha=3.5$ corresponds to the minimal parameter value such that the innovation function is linear in $\mu$.

 \section{Representation of interacting  groups of interest with multiple strategies}

\subsection{Mathematical structure in the KTAP framework }

Following \cite{Bpreprint} we define some hallmarks as paradigms of our modelling approach:

\begin{itemize}

\item  The   society is partitioned into {\it functional subsystems}, representing groups of interests.

\item In each functional subsystem individuals are called {\it active particles} and share a common strategy represented by an  {\it activity variable}.

\item The activity variable is heterogeneously distributed within each functional subsystem attaining a range of values going from a "worst" one to a "best" one.

\item  The state of each functional subsystem is described by a probability distribution on the activity variable.

\item
An evolution equation for the probability distribution on each functional subsystem is obtained by a balance of particles in the elementary volume of the microstates,  the inflows and outflows being related to the defined interactions.

\item We assume binary interactions between active particles belonging to the same fuctional subsystems   as well as interactions of active particles with macroscopic quantities characterizing other functional subsystems, characterizing a {\it stream effect}.

\item Interactions are modeled as {\it stochastic games} in which the payoff is given in probability and are, in general, nonlinearly additive in the sense that they may depend  on the probability distribution itself.

\end{itemize}

The microstate of each functional subsystem is characterized by a bivariate activity variable with two components
\begin{equation}
\{(u_i,\nu_r ),\quad i=0,\dots,I;\quad r=0,\dots,R\}.
\end{equation}
Both components attain a maximal value, so that they can be   normalized with respect to their respective maximal value. The domain of the activity variable is $D_u\times D_\nu$ where
$$\displaystyle{D_u=\{u_i=\frac i{I},\:i=0,\dots, I\}\subset[0,1],\: D_\nu=\{\nu_r=\frac r{R},\:r=0,\dots, R\}\subset[0,1]}.$$

A   probability mass function is defined on the microstate for each functional subsystem
\begin{equation}
f_{ir}^s(t)=f^s(t,u_i,\nu_r)\,:\,[0,1]\times D_u\times D_\nu\:\to\:[0,1],\quad s=1,\dots,S
\end{equation}
representing the number of active particles that at time $t$ express the strategy $(  u_i, \nu_r)$ in the subsystem $s$; time has been normalized with respect to a maximal value $T_{max}$ assumed to exist. Moreover, it has been assumed that the number of active particles in each subsystem remains constant during the evolution, allowing to normalize the distribution with respect to it, so that the following  property   applies
$$\sum_{i=0}^I \sum_{r=0}^R f_{ir}^s(t)=1,\quad \forall  s=1,\dots,S,\:\forall t\in[0,1].$$
The statistical moments of the probability mass function on each functional subsystem allow us to recover the macroscopic quantities (observables) related to each functional subsystem. The first-order moments related to the marginal probabilities on each functional subsystem are
\begin{equation}
\mathbb E^s_\nu(t )=\sum_{i=0}^I\sum_{r=0}^R\nu_r  f_{ir}^s(t),
\end{equation}
\begin{equation}
 \mathbb E^s_ u(t)=\sum_{r=0}^R\sum_{i=0}^I u_i  f_{ir}^s(t),
\end{equation}
${\small s=1,\dots,S}$.

The evolution equation   for each functional subsystem is obtained  by a balance between the inlet and the outlet of active particles in the elementary volume of the microstate,
\begin{equation}\label{generale}
\displaystyle{\frac {d} {dt} f_{ir}^s(t)   =J_s[\mathbf  f^s]+ \mathcal J_s[\mathbf f^j]},
\end{equation}
with the initial condition $(\mathbf f^s)_0=\mathbf f ^s(0)$,  $ s=1,\dots,S$  and $ j\in\{0,\dots,s-1,s+1,\dots S\}$; moreover we used the notation $\displaystyle{\mathbf f^s=\{f_{ir}^s\}, \:i=0,\dots, I,\, r=0,\dots, R }$.
 Interactions are modeled as game-type with stochastic payoffs that are specific for each application and determine the explicit expressions of the following two terms in each subsystem,
\begin{itemize}
\item $J_s[\mathbf  f^s]\:(\footnotesize{  s=0,\dots,S})$ - accounting for the net flow of active particles due to binary interactions among active particles belonging to the same $s$-th functional subsystem; it may depend on the probability mass function of the $s$-th subsystem through its first moments.

\item$\mathcal J_s[\mathbf f^j]\:({ \small s=0,\dots,S},\quad {\small j\in\{0,\dots,s-1,s+1,\dots S\}})$ - accounting for the net flow of active particles in the $s$-th subsystem due to the influence of the the first moments of the $j$-th subsystem.
\end{itemize}

In the following we will derive the explicit expressions for the above introduce flows.

\subsection{Linking economic development and political perspectives}

In consonance with the specific features of AR-model, let us consider a population of individuals subdivided into three functional subsystems: incumbent {\it ruler}, {\it citizens} and a political {\it competing group}, clustered according to   specific strategies that they express in the analyzed competition. In particular, the ruler are characterized by their {\it propensity to innovate} where innovation has the meaning of introduction of technological innovation in a broad sense, as explained in the previous section. Moreover the ruler are characterized by their {\it political power}. The citizens express their {\it wealth} and their {\it political opinion} on the ruler. The competing group is characterized by their {\it wealth} and  their {\it political power}.

One may summarize the   dynamic that we are going  to model as in the following. The propensity to innovate of the incumbent ruler is in a direct relationship with their political power (more political power, more propensity to innovate) and this propensity determines whether innovation is introduced or not in the society. Whenever innovation is introduced in the society, a positive increment in the citizens wealth may be probabilistically  obtained  as well as in the competing group wealth. Citizens political opinion is assumed to be driven only by their wealth and citizens opinion directly influences the political power of the rulers. Analogously,  the political power of the competing group is assumed to depend only on their wealth and it has a negative return on the political power of the ruler. An intuitive graphical representation of this   socio-economic system is given below, where the minus signs and the red colors represent symbolically  negative returns.
 
\begin{figure}[htbp]
\centering{\includegraphics[height=65mm, width=\textwidth]{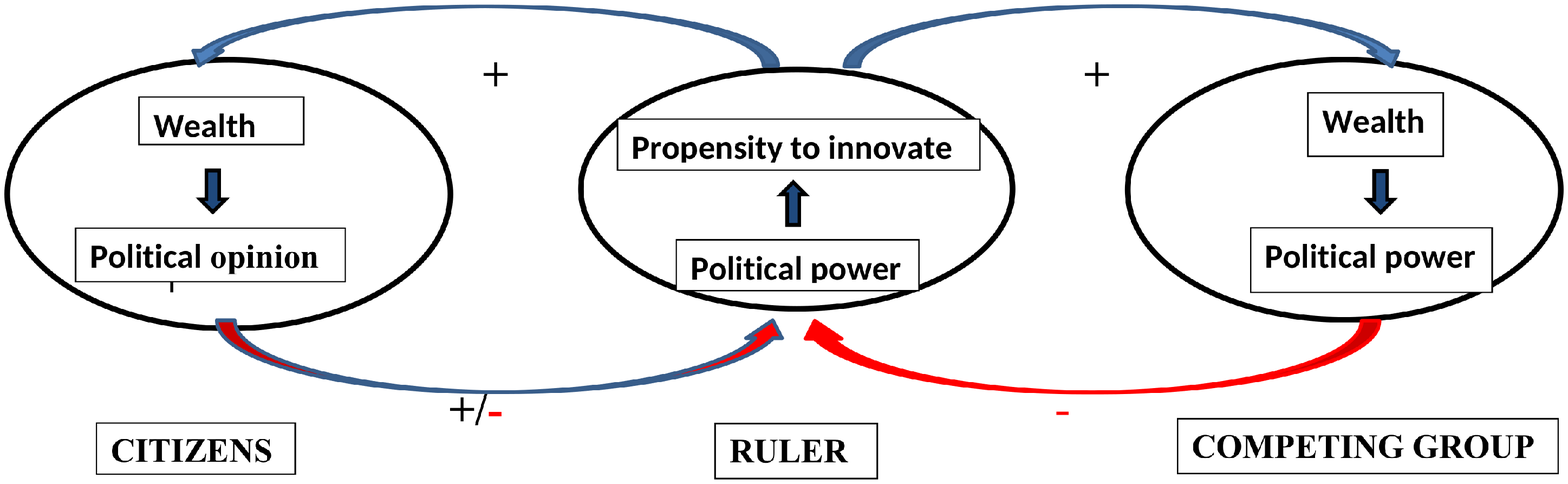}}

\end{figure}


To explictly model the phenomenology above sketched one has to introduce the {\it transition probabilities}, which characterize the probability that has a candidate particle to switch to a different value of activity variable due to interactions. Transition probabilities  are of two types: 
\begin{itemize}
\item $D^{kq}_{hp}(s)(\nu_p\to\nu_r)$  - modeling the probability for a candidate particle with microstate $(u_h,\nu_p)$ of the $s$-th subsystem to  change  its second component   of the activity variable $(\nu_p \to \nu_r)$   due to an interaction with a field particle with microstate $(u_k,\nu_q)$ of the same subsystem.
\end{itemize}
 \begin{remark} By encoding the phenomenology above described of the three interacting groups of interest, it is assumed that the first component of the activity variable of the candidate particle  do not change due to binary interactions among active particles of the same subsystem.\end{remark}
\begin{itemize}
\item $\mathcal B^j_h(s)(u_h \to u_i)$ - modeling the probability that has a candidate particle  of the $s$-th subsystem to change   its first component   of the activity variable $(u_h \to u_i)$ due to the influence of the first-order moments of the $j$-th subsystem ($s=1,2,3,\,j=1,2,3$ with $j\neq s$).\end{itemize}
  \begin{remark} By encoding the phenomenology above described of the three interacting groups of interest, it is assumed that the second component of the activity variable of the candidate particle  do not change due to the influence of  subsystems different with respect to the one to which the particle belongs.\end{remark}

Each subsystem representing {\it ruler}, {\it citizens} and {\it competing group} respectively is   indexed as explained in the following; moreover the activity variable for each of them and the interactions are modeled.

\begin{enumerate}
\vspace{0.4cm}
\item {\bf Subsystem 1. \hspace{.3cm}
Ruler}

{\it Microstate}: $\left\{\begin{array}{l}    \mathrm{political\: power} \hspace{1.8cm}( u_i,\,i=1,\dots, I)\\
\mathrm{propensity\: to\: innovate}\quad ( \nu_r,\,r=0,\dots, R)\end{array}\right.$

\vspace{0.2cm}
{\it Interactions within the subsystem}: \newline
It is assumed that the propensity to innovate of the ruler is influenced by their political power. In particular, 
if the ruler have a high mean  political power,   their propensity to introduce innovation in the society may rise, in probability. The opposite is assumed if the ruler have a low mean political power.

The above sketched phenomenology is encoded in the following transition probabilities which make use of uniform probability mass functions:

\begin{itemize}
\item if $\mathbb E^1_u\geq \frac 1{2},$  
\begin{equation} \displaystyle{\forall k,q,\quad D_{hp}^{kq}(1)(\nu_p\to\nu_r)=\left\{\begin{array}{l}0\hspace{1.3cm}\mathrm{for}\quad r\in\{0,\dots,p-1\} \\\\\frac 1{R-p+1}\quad\mathrm{for}\quad r\in\{p,\dots,R\}
                                                                                  \end{array}\right.}
\end{equation}

\item if $\mathbb E^1_u< \frac 1{2},$  
\begin{equation} \displaystyle{\forall k,q,\quad D_{hp}^{kq}(1)(\nu_p\to\nu_r)=\left\{\begin{array}{l} \frac 1{p}\quad\mathrm{ for}\quad r\in\{0,\dots,p-1\}\\\\
                                                                                  0\hspace{0.5cm}\mathrm{for}\quad r\in\{p,\dots,R\}\end{array}\right.}
\end{equation}
(in both cases above the transition probabilities are independent on the values of $\nu_k$ and $\nu_q$).
\end{itemize}

{\it Influences of the other subsystems}: \newline
It is assumed that the political opinion of the citizens influences the political power of the ruler. In particular,
if the citizens have an high mean political opinion,  the politcal power of the ruler
  may rise, in probability. The opposite is assumed if the citizens  have a low mean political opinion.
  
The above sketched phenomenology is encoded in the following transition probabilities which make use of uniform probability mass functions:

\begin{itemize}
\item if $\mathbb E^2_\nu\geq \frac 1{2},$ 
\begin{equation} \displaystyle{\mathcal B_{h}^2(1)(u_h\to u_i)=\left\{\begin{array}{l}    0\hspace{1.4cm}\mathrm{for}\quad i\in\{0,\dots,h-1\}\\\\\frac 1{I-h+1}\hspace{.6cm}\mathrm{ for}\quad i\in\{h,\dots,I\}
                                                                               \end{array}\right.}
\end{equation}

\item if $\mathbb E^2_\nu< \frac 1{2},$  
\begin{equation} \displaystyle{\mathcal B_{h}^2(1)(u_h\to u_i)=\left\{\begin{array}{l} \frac 1{h}\quad\mathrm{for}\quad i\in\{0,\dots,h-1\}\\\\
                                                                                  0\hspace{.5cm}\mathrm{ for}\quad i\in\{h,\dots,I\}\end{array}\right.}
\end{equation}

\end{itemize}

Actually, it is assumed that the political power of the  competing group influences the political power of the ruler. In particular, if the competing group has a high mean political power, the political power of the ruler may decrease, in probability. If the competing group has a low mean political power, it is assumed that there is no influence on the political power of the citizens.
The above sketched phenomenology is encoded in the following transition probabilities:

\begin{itemize}
\item if $\mathbb E^3_\nu\geq \frac 1{2},$  
\begin{equation} \displaystyle{\mathcal B_{h}^3(1)(u_h\to u_i)=\tilde\gamma \delta_{i,h-1}+(1-\tilde\gamma)\delta_{i,h}}
\end{equation}
 
\item if $\mathbb E^3_\nu<\frac 1{2}$,
\begin{equation} \displaystyle{\mathcal B_{h}^3(1)(u_h\to u_i)=\delta_{h,i}}
\end{equation}
\end{itemize}
In the above equations and hereafter   $\delta_{h,k} $ represents the Kronecker delta.

\item {\bf Subsystem 2. \hspace{.3cm}
Citizens}

{\it Microstate}: $\left\{\begin{array}{l}    \mathrm{wealth} \hspace{2.1cm}( u_i,\,i=0,\dots, I)\\
\mathrm{political \: opinion}\quad ( \nu_r,\,r=0,\dots, R)\end{array}\right.$

\vspace{0.2cm}
{\it Interactions within the subsystem}: \newline
It is assumed that the opinion dynamic of citizens is driven by economic motivations: when two active particles interact,  if the wealth status of the candidate particle is below the one of the field particle, the candidate particle has a probability to acquire the opinion of the field one, following an {\it imitation rule}. If, on the contrary, the wealth status of the candidate particle is above or equal to the one of the field particle, the candidate one will not change his opinion. The following transition probabilities encodes the above described phenomenology:

\begin{itemize}
\item if $u_h<u_k$,  
\begin{equation} \displaystyle{D_{hp}^{kq}(2)(\nu_p\to\nu_r)=\beta\,\delta_{r,q} + (1-\beta)\delta_{r,p}}
\end{equation}

\item if $u_h\geq u_k$,
\begin{equation} \displaystyle{D_{hp}^{kq}(1)(\nu_p\to\nu_r)=\delta_{r,p}}
\end{equation}
where $\beta$ is a parameter of the model.
\end{itemize}

{\it Influences of the other subsystems}: \newline
We model the impact of the introduction of technological innovations on the citizen income by simply reasoning that
if the ruler introduces technological innovation,  the citizens may rise their wealth status. 
 In particular, if the ruler an high mean propensity to innovate, the wealth of the citizens may rise, in probability. If the ruler have a low mean propensity to innovate, it is assumed that there is no influence on the wealth of the citizens.
The above sketched phenomenology is encoded in the following transition probabilities:

We model this effect by assuming that
the  ruler may have a positive   return on the wealth of the citizens depending on the mean value of their propensity to innovate ($\mathbb E^1_\nu$). The transition probabilities encoding the above described phenomenology are the following,

\begin{itemize}
\item if $\mathbb E^1_\nu\geq \frac 1{2}$,  
\begin{equation}
\label{rents} \displaystyle{\mathcal B_{h}^1(2)(u_h\to u_i)=\tilde\alpha\, \delta_{i,h+1}+ (1-\tilde\alpha)\delta_{i,h }}
\end{equation}

  \item if $\mathbb E^1_\nu< \frac 1{2}$,  
\begin{equation} \displaystyle{\mathcal B_{h}^1(2)(u_h\to u_i)= \delta_{h,i}}
\end{equation}

\end{itemize}

\item {\bf Subsystem 3. \hspace{.3cm}
Competing group}

{\it Microstate}: $\left\{\begin{array}{l}    \mathrm{wealth} \hspace{1.9cm}( u_i,\,i=0,\dots, I)\\
\mathrm{political\:power}\quad ( \nu_r,\,r=0,\dots, R)\end{array}\right.$

\vspace{0.2cm}
{\it Interactions within the subsystem}: \newline
It is assumed that the political power of the competing group is influenced by their wealth.  In particular, 
if the competing group has a high mean  wealth,   their political power   may rise, in probability. The opposite is assumed if the competing group have a low mean wealth.

The above sketched phenomenology is encoded in the following transition probabilities which make use of uniform probability mass functions:

\begin{itemize}
\item if $\mathbb E^3_u\geq \frac 1{2}$,  
\begin{equation} \displaystyle{\forall k,q,\quad D_{hp}^{kq}(3)(\nu_p\to\nu_r)=\left\{\begin{array}{l}  0\hspace{1.3cm}\mathrm{ for}\quad r\in\{0,\dots, p-1\}\\\\ \frac 1{R-p+1}\quad\mathrm{ for}\quad r\in\{p,\dots,R\}
                                                                                \end{array}\right.}
\end{equation}

\item if $\mathbb E^3_u< \frac 1{2}$,  
\begin{equation} \displaystyle{\forall k,q,\quad D_{hp}^{kq}(3)(\nu_p\to\nu_r)=\left\{\begin{array}{l} \frac 1{p}\quad\mathrm{ for}\quad r\in\{0,\dots, p-1\}\\\\
                                                                                  0\hspace{0.5cm}\mathrm{ for}\quad r\in\{p,\dots,R\}\end{array}\right.}
\end{equation}
(in both cases above the transition probabilities are independent on the values of $\nu_k$ and $\nu_q$).
\end{itemize}

{\it Influence of the other subsystems}: \newline
We model the impact of the introduction of technological innovations on the competing group income too by simply reasoning, as in the case of the citizens, that
if the ruler introduces technological innovation,  the competing group  may rise their wealth status, analogously to the case of the citizens. 
 In particular, if the ruler a high mean propensity to innovate, the wealth of the competing group may rise, in probability. If the ruler have a low mean propensity to innovate, it is assumed that there is no influence on the wealth of the competing group.
The above sketched phenomenology is encoded in the following transition probabilities:

\begin{itemize}
\item if $\mathbb E^1_\nu\geq \frac 1{2}$ 
\begin{equation} \displaystyle{\mathcal B_{h}^1(3)(u_h\to u_i)=\left\{\begin{array}{l}  0\hspace{1.2cm}\mathrm{ for}\quad i\in\{0,\dots,h-1\}\\\\\frac 1{I-h+1}\quad\mathrm{ for}\quad i\in\{h,\dots,I\}
                                                                                 \end{array}\right.}
\end{equation}

  \item if $\mathbb E^1_\nu< \frac 1{2}$ 
\begin{equation} \displaystyle{\mathcal B_{h}^1(3)(u_h\to u_i)= \delta_{h,i}}
\end{equation}

\end{itemize}

\end{enumerate}

In order to derive the explicit expressions for the flows of eq.(\ref{generale}) and then the kinetic model with stochastic game-type interactions, we need also to introduce the frequency of binary  interactions of the candidate particle with the field particle (regarding the dynamic within each subsystem) or with first-order moments of other subsystems. The frequency of interactions are also called {\it encounter rates}; in particular:
\begin{itemize}
\item $\eta_s$ - encounter rate within the $s$-th subsystem, i.e. the frequency of interactions of the candidate particle with the field particle within the same subsystem ($s=1,2,3$). We adopt a constant rate of interactions within each subsystem.
\item $\mu_s^j$ - encounter rate among the subsystems, i.e. the frequency of interactions of the candidate particle of the $s$-th subsystem with the first order moments of the $j$-th subsystem ($s,\,j=1,2,3,\,j\neq s$). We adopt a constant rate of interactions between each couple of interacting functional subsystems.
\end{itemize}

We are ready now to derive the explicit forms of the
   flows of eq.(\ref{generale}) by balancing the inflow and the outflow of active particles in the elementary volume of the microstate:

\begin{equation} 
\displaystyle{ J_s[\mathbf  f^s]= 
\sum_{h,k=0}^I\,\sum_{p,q=0}^R \eta_s\, D^{kq}_{hp}(s)(\nu_p\to\nu_r)f^s_{hp} \:f^s_{kq}}
 \displaystyle{ -f_{ir}^s  \sum_{k=0}^I\sum_{q=0}^R \eta_s\,  f_{kq}^s,} 
\end{equation}

\begin{equation}
\displaystyle{  \mathcal J_s[ \mathbf  f^j]=\sum_{ \begin{array}{l} j=1\\ j\neq s\end{array}}^3 \Big\{ \sum_{h=0}^I\sum_{p=0}^R \Big[
\mu^j_s\,\mathcal B^j_h(s)(u_h \to u_i )f^s_{hp}  \Big]
- \mu^j_s f_{ir}^s\Big\},}
\end{equation}
for $ s=1,2,3.$

\begin{remark} The parameter $\mu$, quantifying the inverse  of the  political competition can be compared to  
   the first moment with respect to the political power of  the  competing group, then it is given by $\displaystyle{  \mathbb E^3_\nu}$. \end{remark}

\begin{remark} 

The parameter $\alpha$, quantifying  the effect of the introduction of technological innovation in AR-model, can be compared to  the $\tilde\alpha$ present in eq.(\ref{rents}) which quantifies the effect of the introduction of technological innovation by the ruler on the wealth of citizens .

 \end{remark}

\begin{remark} 
The parameter  $\gamma$ of AR-model, quantifying  the 'erosion' of the political power of the incumbent ruler due to the introduction of new technologies in the society,  can be compared to the parameter $\tilde\gamma$, which quantifies the effect of the   political power of the competing group on the political power of the ruler.
\end{remark}

\section{Numerical solutions and critical analysis }

In the numerical solutions the parameters of the transition probabilities take the values of  Tab. \ref{tabella}.
\begin{table}[h]
\begin{tabular}{cp{8cm}cp{8cm}cp{8cm} }
{\em Parameter}         &       {\em  Meaning   }                                                                   &         {\em Value}\\ \hline
$\tilde\alpha$          &scales the positive return on the citizen wealth                                 & 0.1\\  \hline
$\beta$                    &citizen susceptibility to change opinion                                           &0.3\\  \hline
$\tilde\gamma$         &scales the negative return on the political power of  the competing group& 0.9\\  \hline

\end{tabular} \caption {Parameters involved in the transition probabilities.}\label{tabella}
\end{table}

 All the encounter rates are settled as 1.

\paragraph{{\bf Case Study I - Strong ruler with weak political competing group: nonmonotonous relationship between political competion and propensity to innovate of the ruler. }}

\begin{figure}[h!]
\centering
\includegraphics[height=4cm,width=13cm]{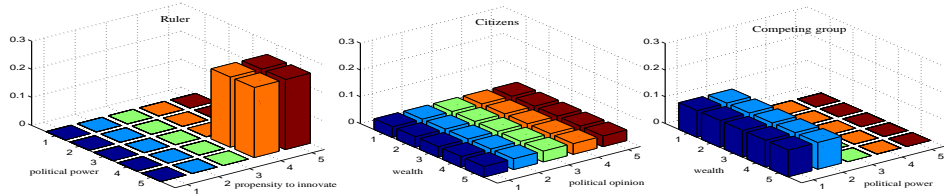}
 
\caption{Initial conditions on the three functional subsystems.}
\label{caseI_initial}
\end{figure}

We assume as initial condition a strong ruler Fig.(\ref{caseI_initial}$)_a$ and a weak competing group Fig.(\ref{caseI_initial}$)_c$, whilst the wealth and political opinion of the citizens are homogeneously distributed (\ref{caseI_initial}$)_b$.

\begin{figure}[h!]
\centering
\includegraphics[height=4cm,width=12cm]{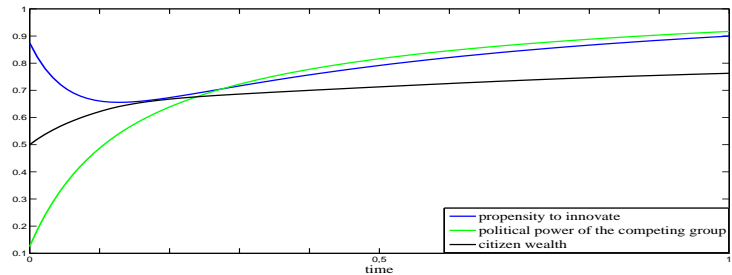}

\caption{ Evolutions of the propensity of the ruler to innovate ($\mathbb E^1_\nu$), citizen wealth ($\mathbb E^2_u$) and political power of the competing group ($\mathbb E^3_\nu$) in  a society with strong ruler and weak opposition}
 \label{caseI_meanvalues}
\end{figure}

In Fig.(\ref{caseI_meanvalues}) it is reported the evolution of the first order moments referred to the propensity to innovate of the ruler (blue), the political power of the competing group (green) and the citizen wealth (black).

\begin{figure}[h!]

\quad\includegraphics[height=4cm,width=12cm]{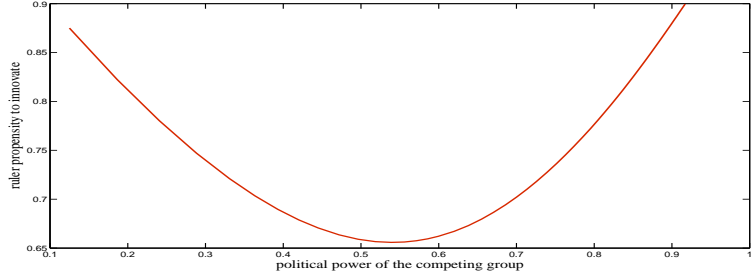}
\caption{  Propensity to innovate of the ruler ($\mathbb E^1_\nu$)  vs. political power of the competing group ($\mathbb E^3_\nu$) in a society with strong ruler and weak opposition.}
  \label{nonmonotonous}
\end{figure}

  Fig.(\ref{nonmonotonous})  shows the nonmonotonous relationship between the propensity to innovate of the ruler and the political power of the competing group.

The general observed trends are  conserved also if one changes  the initial distribution of wealth and political opinion of the citizens.

\newpage
 
\paragraph{ {\bf Case study II - Strong ruler with strong political opposition. }}

We show emergent behaviors when a ruler with strong political power acts in a society with a high level of political competition.  The   wealth and political opinion of the citizen are taken as homogeneously distributed; however the general trends are conserved as the initial distribution on the citizens changes.

\begin{figure}[h!]
\centering
\includegraphics[height=4cm,width=12cm]{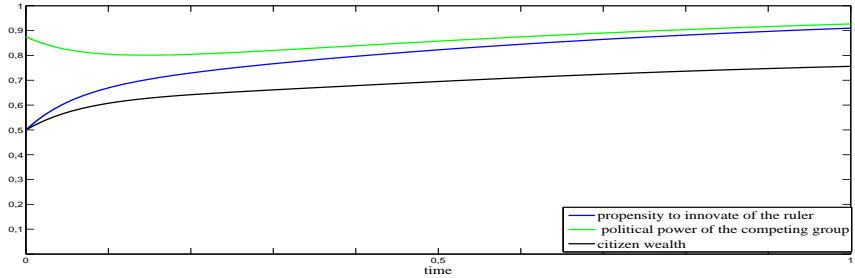}

\caption{Evolutions of the propensity of the ruler to innovate ($\mathbb E^1_\nu$), citizen wealth ($\mathbb E^2_u$) and political power of the competing group ($\mathbb E^3_\nu$) in  a society with a strong rule and a strong political competition.}
\label{emergentIIa} 
\end{figure}

\begin{figure}[h!]
\centering
\includegraphics[height=4cm,width=12cm]{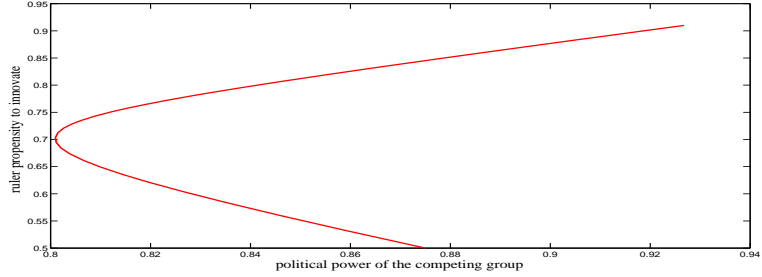}
\caption{  Propensity to innovate of the ruler ($\mathbb E^1_\nu$)  vs. the political power of the competing group ($\mathbb E^3_\nu$) in a society with strong ruler and strong opposition.}
\label{emergentIIb} 
\end{figure}

The tendency to introduce innovation of the ruler increases (Fig.\ref{emergentIIa}).
When analysing the propensity to innovate of the ruler    vs. the political power of the competing group it is observed that the propensity to innovate is always increasing whilst the political power of the competing group is decreasing in a first interval and then it is increasing. So there is a nonmonotonous relationship of the political power of the competing group vs. the propensity to innovate of the ruler.

The general observed trends are  conserved also if one changes  the initial distribution of wealth and political opinion of the citizens.

\paragraph{ {\bf  Case study III - Weak ruler with strong opposition. }}

We show emergent behaviors when a ruler with weak political power acts in a society with a strong  level of political competition.

\begin{figure}[h!]
 
\centering
 \includegraphics[height=4cm,width=12cm]{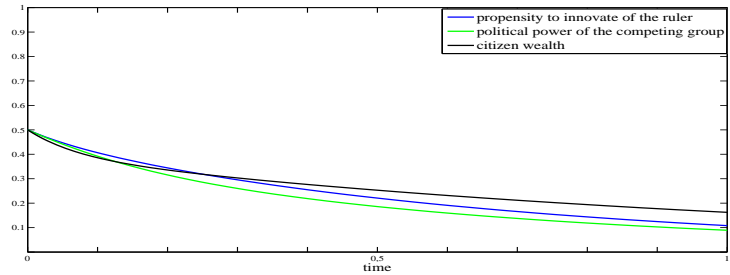} 
\caption{Evolutions of the propensity of the ruler to innovate ($\mathbb E^1_\nu$), citizen wealth ($\mathbb E^2_u$) and political power of the competing group ($\mathbb E^3_\nu$) in  a society with {\bf  weak ruler and strong opposition}}
\label{emergentIII_a}
\end{figure}

\begin{figure}[h!]
\centering
 \includegraphics[height=4cm,width=12cm]{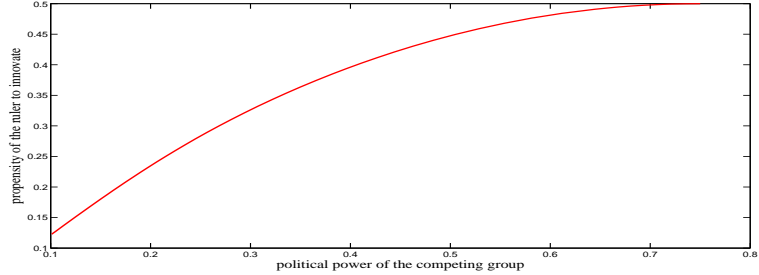} 
\caption{Propensity to innovate of the ruler ($\mathbb E^1_\nu$)  vs. political power of the competing group ($\mathbb E^3_\nu$) in  a society with {\bf  weak ruler and strong opposition}}
\label{emergentIII_b}
\end{figure}

In this situation the tendency of the ruler to introduce innovation decreases (Fig.\ref{emergentIII_a}) and the relationship between the propensity of the ruler to innovate and the political power of the competing group is monotonous (Fig.\ref{emergentIII_b}).

The general observed trends are  conserved also if one changes  the initial distribution of wealth and political opinion of the citizens.

\newpage

\paragraph{ {\bf Case study IV - 
Balanced political power of the ruler and the competing group in a poor society  and in a rich society. }}

In this case study we consider the case of a society with medium political power both for the ruler and the competing group.
We show that  the time evolution of the mean value of propensity to introduce innovation of the ruler and political power of the competing group are dependent on the initial distribution of citizen wealth.

\begin{figure}[h!]
 
\centering
\includegraphics[height=3cm,width=6.5cm]{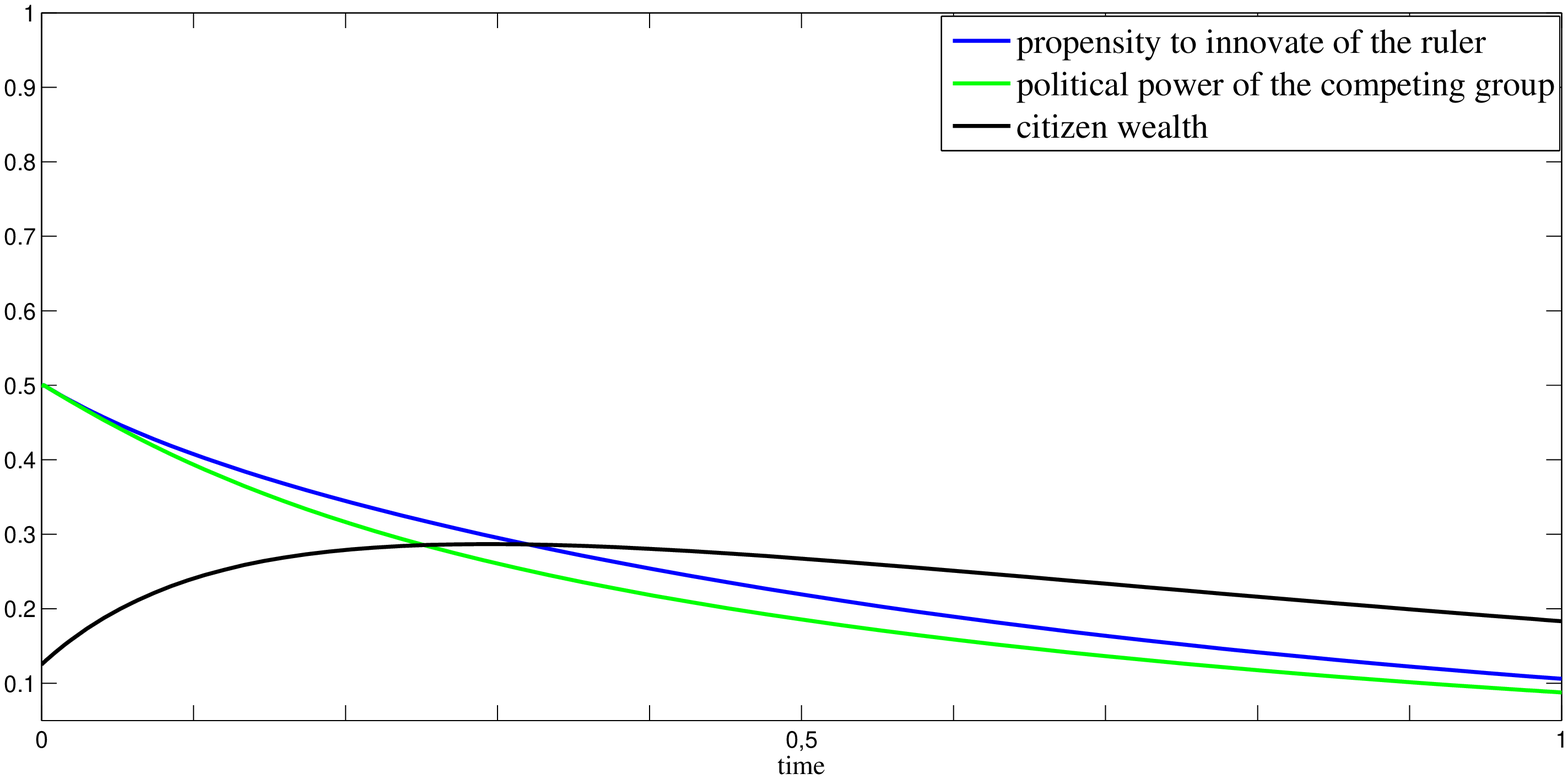}\quad\includegraphics[height=3cm,width=6.5cm]{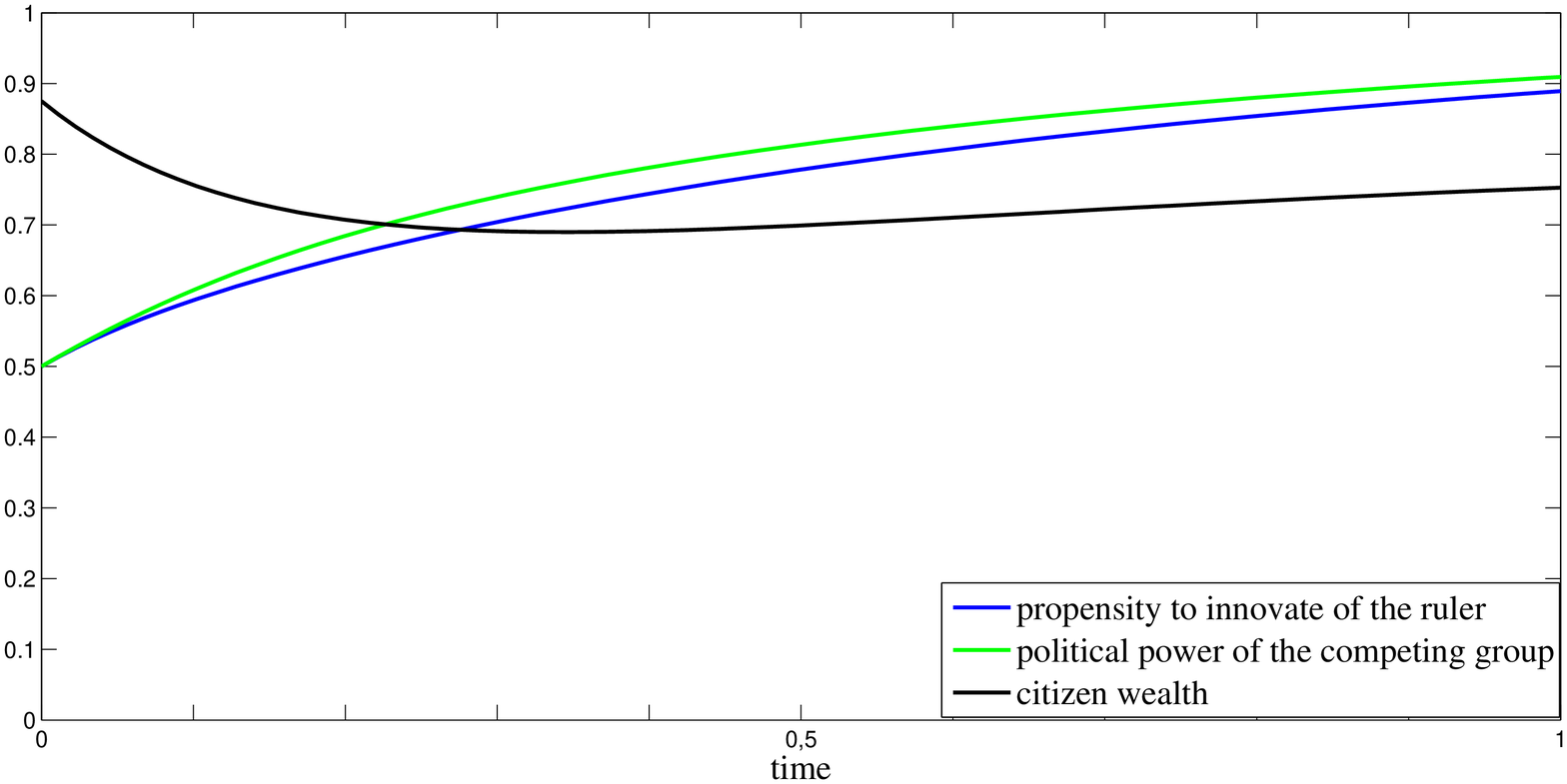} 
\caption{Case study IV: Evolutions of the propensity of the ruler to innovate ($\mathbb E^1_\nu$), citizen wealth ($\mathbb E^2_u$) and political power of the competing group ($\mathbb E^3_\nu$) in  a society with  {\bf 
balanced political power of the ruler and the competing group} in a poor society (left) and in a rich society (right).}
\label{emergentIV}
\end{figure}

\begin{figure}[h!]
 
\centering
\includegraphics[height=4cm,width=12cm]{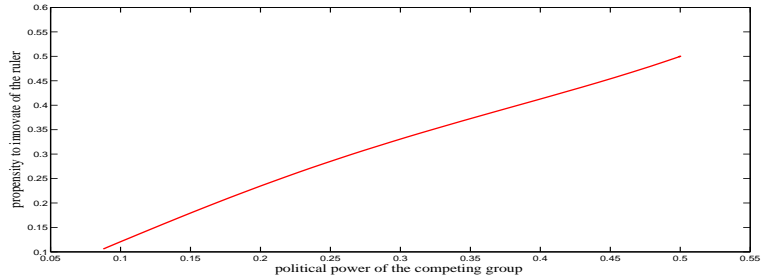}
\caption{Propensity of the ruler to innovate vs. political power of the competing group in a poor society.}
\label{emergentIV_b}
\end{figure}

\begin{figure}[h!]
 
\centering
\includegraphics[height=4cm,width=12cm]{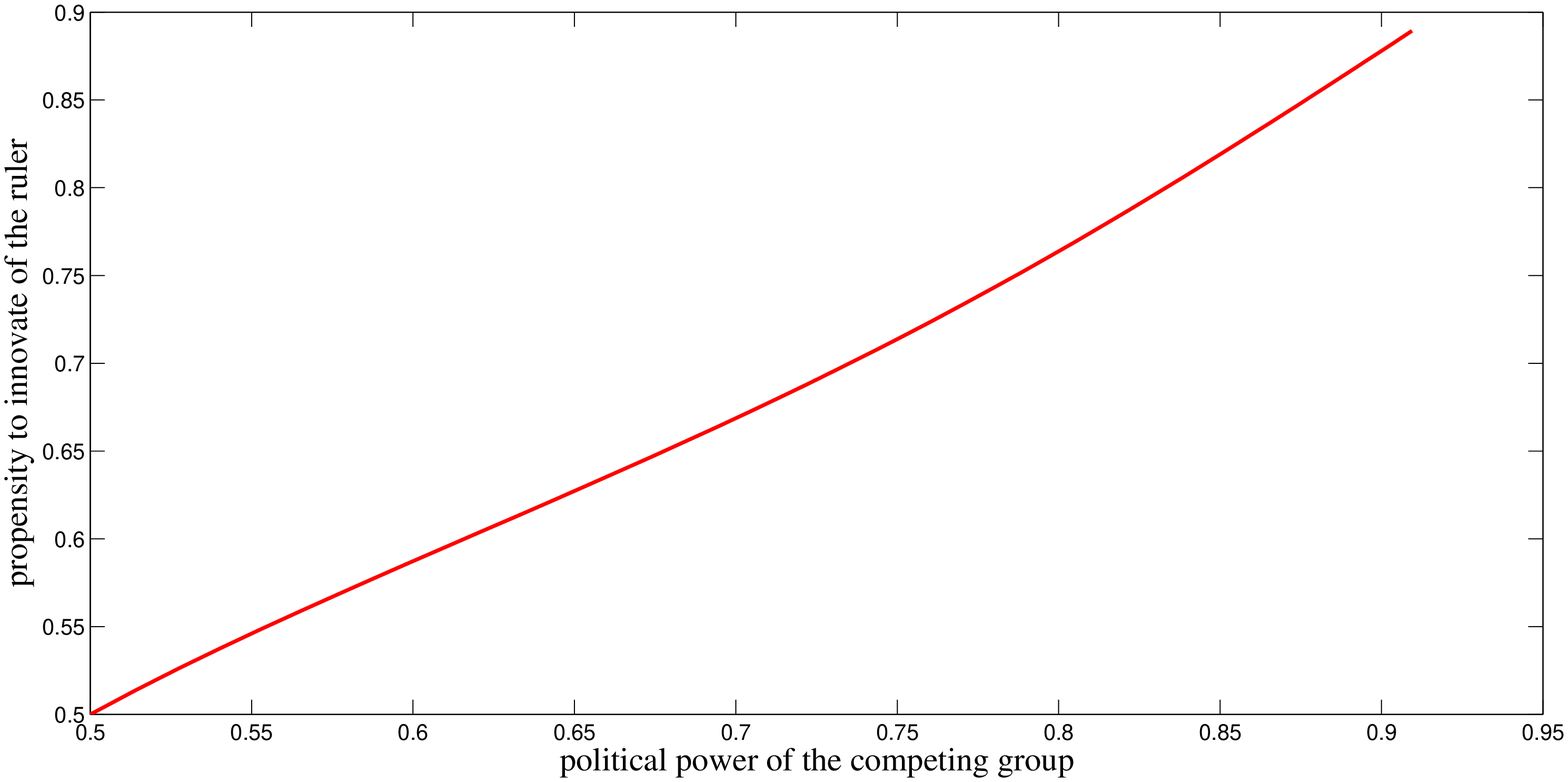}
\caption{Propensity of the ruler to innovate vs. political power of the competing group in a wealthy society.}
\label{emergentIV_b1}
\end{figure}

One observes that when the political power of the ruler and that of the competing group are balanced,  the propensity to innovate of the ruler tends to decrease in the case of a poor society (Fig.\ref{emergentIV} on the left) whilst it tends to increase in a wealthy society (Fig.\ref{emergentIV} on the right). In both cases a monotonous relationship between the propensity of the ruler to innovate and the political power of the competing group is observed (Fig.\ref{emergentIV_b} and Fig.\ref{emergentIV_b1}, respectively).

-

\paragraph {\bf Case study V: Sensitivity to the initial conditions of the ratios between first order moments.}

Let us finally consider the time evolution of the ratios between some of the first-order moments.

 Let us define the following quantities:
$$F(t) = \frac{\mathbb{E}_u^1}{\mathbb{E}_\nu^3}=\frac{\mathrm{ ruler\: political \: power}}{\mathrm {competing\: political \: power}}$$
and the propensity to innovate of the ruler and the political power of the competing group

$$G(t) = \frac{\mathbb{E}_\nu^1}{\mathbb{E}_\nu^3}=\frac{\mathrm{ ruler\: propensity\:to\:innovate}}{\mathrm {competing\: political \: power}}$$
  with the scope to explore the sensitivity to the initial conditions for both cases.

Again, simulations are performed with parameters $\tilde{\alpha} = 0.1$, $\beta = 0.3$ and $\tilde{\gamma} = 0.9$, and we seek for emergent behaviours while varying initial conditions. In particular, different values of $F(0)$ and $G(0)$ are obtained by initially clustering rulers and political opponents in different activity groups, while citizens are assumed to have a uniform initial distribution in all cases.

\begin{figure}[h]
\centering
\includegraphics[height=3cm,width=6cm]{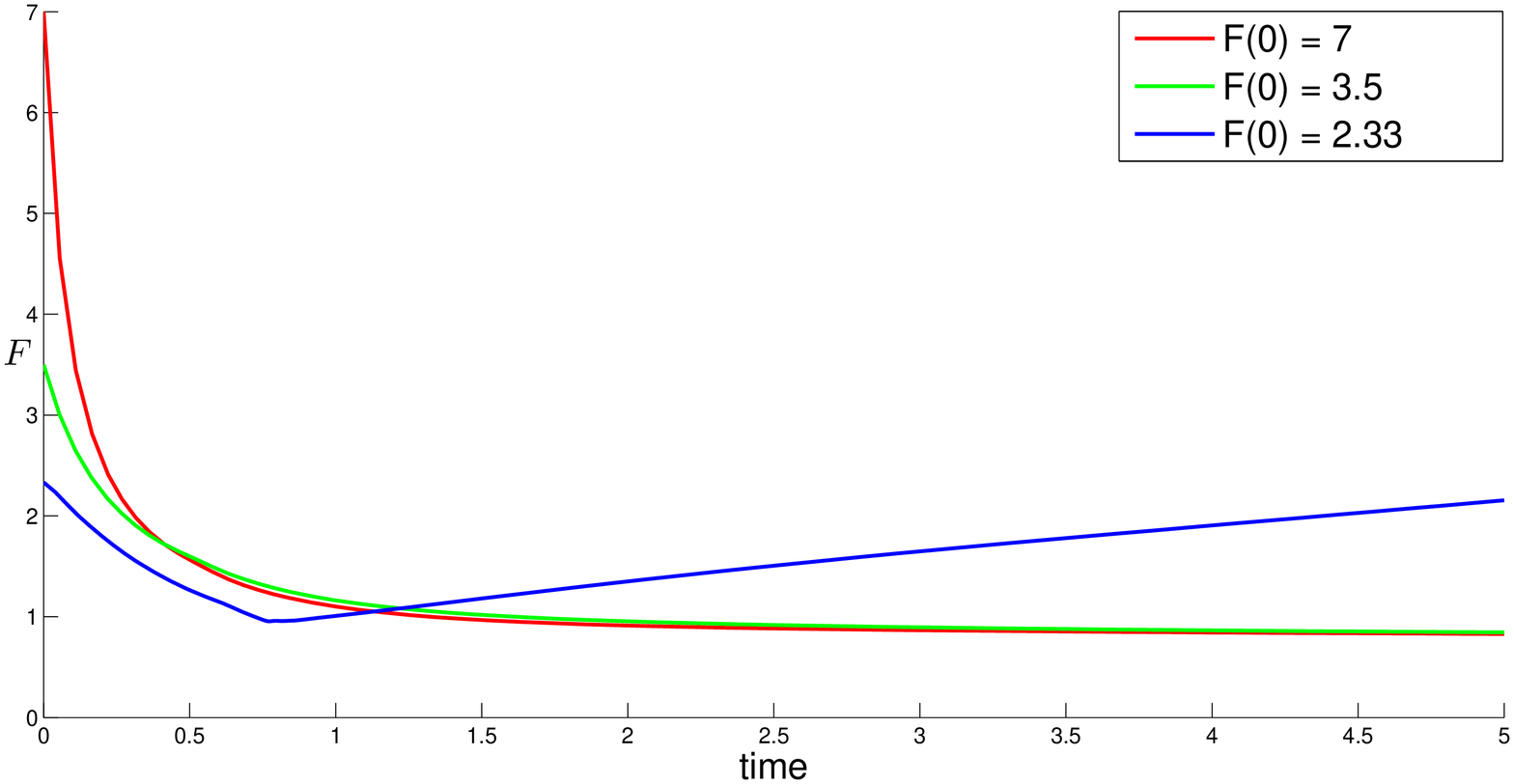}
\quad\includegraphics[height=3cm,width=6cm]{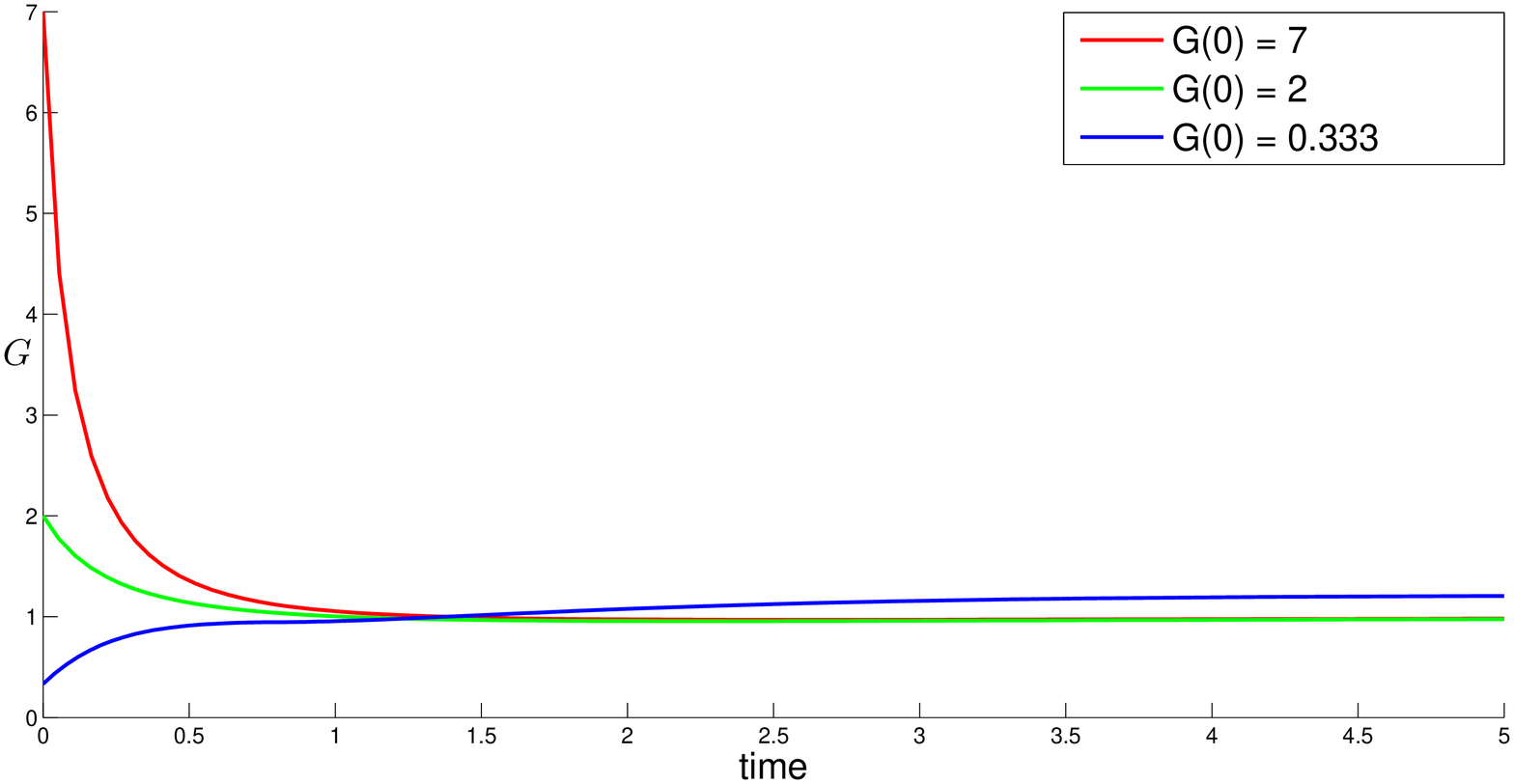}
\caption{Evolution of the quantities $F(t)$ (left) and $G(t)$ (right)}
\label{emergentV}
\end{figure}

Results are shown in Fig. \ref{emergentV}. We can see 
that even for opposite initial conditions, $G(t)$ shows a trend to an asymptotic value near to $1$, that is the propensity to innovate of the ruler tends to be equilibrated by the competing group's political power. An analogous result is observed for $F(t)$, that compares both groups' political power, that shows a trend to a value near to $0.8$. These results show that when the ruler are politically weaker than the competing group, they are able to surpass it for some time but, after a while, their political power decreases again. This behaviour looks very interesting and a suitable explanation should be found.

\section{Conclusions and perspectives}  
The differences in economic policies and political institutions is a central issue in political economy, tending to clarify the mechanics of cross-country income differences. Some evidences from the US can be found in \cite{economy1}.
 Acemoglu and Robinson   argued that the incumbent ruler is in power being selected as a good match for the actual 'environment' but, due to the introduction of innovations, the environment changes decreasing the ruler's advantage and inducing a mechanism that may led to 'blocking' of the political reforms.

 In the present paper we propose a kinetic model where interactions are modeled as stochastic games; three groups characterized by different strategies (ruler, citizens and a political competing group) evolve by rules internal within each group and, at the same time, being influenced by the dynamics and evolutions of the other groups.
A set of numerical solutions is obtained in order to analyse the conditions determining the nonmonotonous relationship between the propensity of the ruler to innovate and the political competition of the society conjectured by AR-model.

An interesting development consists in considering an external action, which is a development of the basic AR-model in \cite{acemoglu} and is represented by the effect of an external threat on the society. In \cite{acemoglu} it is showed that   the blocking region, resulting in a parameter range of values  is, in that case, significatively reduced.

\paragraph

\end{document}